# Privacy Ethics Alignment in AI: A Stakeholder-Centric Based Framework for Ethical AI

Ankur Barthwal
*Computer Science Department*
*Vancouver Island University*
Nanaimo, Canada
ankur.barthwal@viu.ca

Molly Campbell
*Computer Science Department*
*Vancouver Island University*
Nanaimo, Canada
molly.campbell@viu.ca

Ajay Kumar Shrestha
*Computer Science Department*
*Vancouver Island University*
Nanaimo, Canada
ajay.shrestha@viu.ca

***Abstract*****—** The increasing integration of artificial intelligence (AI) in digital ecosystems has reshaped privacy dynamics, particularly for young digital citizens navigating data-driven environments. This study explores evolving privacy concerns across three key stakeholder groups—young digital citizens, parents/educators, and AI professionals—and assesses differences in data ownership, trust, transparency, parental mediation, education, and risk–benefit perceptions. Employing a grounded theory methodology, this research synthesizes insights from key participants through structured surveys, qualitative interviews, and focus groups to identify distinct privacy expectations. Young digital citizens emphasized autonomy and digital agency, while parents and educators prioritized oversight and AI literacy. AI professionals focused on balancing ethical design with system performance. The analysis revealed significant gaps in transparency and digital literacy, underscoring the need for inclusive, stakeholder-driven privacy frameworks. Drawing on comparative thematic analysis, this study introduces the Privacy–Ethics Alignment in AI (PEA-AI) model, which conceptualizes privacy decision-making as a dynamic negotiation among stakeholders. By aligning empirical findings with governance implications, this research provides a scalable foundation for adaptive, youth-centered AI privacy governance.

## I. Introduction

Artificial Intelligence (AI) has become a fundamental part of everyday digital interactions, profoundly influencing how young digital citizens engage with technology. These individuals are often defined as those who have grown up in a digital environment where AI-driven applications shape their online behaviors [1], [2]. They regularly interact with AI-based services such as social media platforms, virtual assistants, and educational tools [3]. While these technologies offer benefits, such as personalized experiences and enhanced engagement, they also raise critical privacy concerns. AI systems typically rely on extensive data collection, automated decision-making, and complex, often unclear algorithms, leading to concerns about data ownership, user control, trust, and transparency [4].

The growing dependence on AI-driven systems has heightened concerns about privacy management, algorithmic accountability, and the ethical use of data. Research indicates that young users often lack a clear understanding of how their personal data is processed, shared, and monetized [5]. Moreover, many AI technologies function as "black-box" systems, which limits transparency making it difficult for users to make informed privacy decisions [6]. Factors such as parental guidance, regulatory frameworks, and digital literacy initiatives play a crucial role in how young digital citizens navigate privacy risks [7]. However, much of the existing research focuses on adult users or general privacy concerns, leaving a significant gap in understanding how youth-specific factors influence privacy management behaviors in AI ecosystems [8]. In addition, current literature often overlooks the perspectives of varying stakeholders, such as educators, policymakers, and AI developers, whose insight could provide a more comprehensive understanding of the privacy challenges facing young digital citizens. The lack of diverse stakeholder perspectives limits the development of ethical and stakeholder-driven policies and practices that prioritize the privacy and well-being of young users.

To address these gaps, this study employs the Privacy-Ethics Alignment in AI (PEA-AI) model to explore how young digital citizens and key stakeholders, including parents, educators, and AI professionals, perceive and manage privacy in AI-driven environments. The PEA-AI Model emerges from an inductive data-driven approach to understanding how stakeholder perspectives shape ethical AI development. Developed using grounded theory [9], this model analyzes the complex relationship between privacy constructs, stakeholder roles, and ethical AI frameworks. Grounded theory provides a systematic, comparative analysis of privacy concerns, behaviors, and decision-making patterns across different stakeholder groups, allowing for a structured comparison of their perspectives rather than identifying causal relationships. By leveraging empirical data from young digital citizens, parents and/or educators, and AI professionals, the PEA-AI Model provides a negotiation-based framework for privacy governance in AI systems. This approach ensures that variations in privacy expectations, control mechanisms, and transparency demands are examined in alignment with each group's role in the AI ecosystem.

This research focuses on five key constructs influencing privacy management: Data Ownership and Control (DOC) examines perceptions of data ownership, user autonomy, and AI governance mechanisms [1]; Parental Data Sharing (PDS) investigates the role of parental intervention in shaping youth privacy attitudes and AI literacy [4]; Perceived Risks and Benefits (PRB) evaluates how youth balance AI-related risks (privacy breaches, data misuse) against perceived benefits (personalization, efficiency) [5]; Transparency and Trust (TT) explores the relationship between algorithmic explainability, user trust, and privacy decision-making [6]; and Education and Awareness (EA) assesses the impact of AI literacy, privacy education, and digital awareness on privacy management behaviors [8]. Together, these constructs and the PEA-AI Model aim to provide a comprehensive understanding of how diverse stakeholder perspectives can inform ethical privacy governance in AI systems.

This study uses a mixed-methods approach, incorporating quantitative survey data with qualitative insights from open-ended questions, interviews, and focus groups. Participants include young digital citizens (aged 16–19), parents and educators, and AI professionals, providing a comprehensive view of privacy challenges across multiple perspectives. By combining empirical evidence with theoretical insights, this research seeks to contribute to the development of AI privacy policies, ethical AI frameworks, and user-centric design principles.

The rest of this paper is structured as follows: Section II reviews relevant literature. Section III outlines the methodology. Section IV presents key findings. Section V discusses implications, policy recommendations, and future research directions. Section VI concludes the paper by summarizing core insights on AI privacy management for young digital citizens.

## II. OVERVIEW

### A. *Privacy Challenges in AI-Driven Digital Ecosystems*

The rapid expansion of AI-driven technologies has led to unprecedented amounts of data collection and analysis, fundamentally transforming digital interactions and reshaping privacy expectations [1]. Young digital citizens, who regularly engage with AI-powered platforms, encounter unique privacy risks due to the cryptic design of AI systems, lack of user control, and constantly evolving data governance policies [3]. These challenges are further amplified by the socio-technical complexities of AI, where algorithmic decisions impact various aspects of daily life, from social media interaction to personalized educational experiences [7].

While existing research highlights several critical privacy concerns, it often focuses on adults or general user populations, overlooking how youth-specific factors, such as digital literacy, cognitive development, and parental involvement affect privacy management [4]. Additionally, current regulatory and technological frameworks tend to be reactive rather than proactive, lacking a systematic comparison of how different stakeholder groups perceive AI transparency, assess risk, and approach data-sharing behaviours[6]. This gap highlights the need for an integrated, multi-stakeholder approach that addresses comparative privacy concerns across diverse user groups.

While education and awareness have been suggested as essential interventions [8], research indicates that young users often struggle to grasp the intricacies of algorithmic processing and data ownership [7]. The growing commodification of personal data further complicates the issue, as AI-driven platforms operate within a framework of behavioral tracking and predictive analytics, often with minimal user consent or understanding [5].

This study addresses a critical gap by utilizing grounded theory [9] to explore how young digital citizens and key stakeholders (parents, educators, and AI professionals) conceptualize privacy, make trade-offs, and navigate AI-driven ecosystems. Unlike previous studies that focus on privacy attitudes in static digital environments, this research employs a comparative analysis framework to systematically examine privacy expectations, governance concerns, and risk perceptions across stakeholder groups. By addressing privacy concerns within dynamic, algorithmically mediated spaces, the study highlights on the distinct ways in which different stakeholders interpret transparency, data control, and AI ethics. The findings aim to contribute to practical strategies for AI governance, ethical data practices, and privacy education, ensuring that AI systems align with the privacy expectations of young users.

### B. *Conceptualizing Privacy Through Key Constructs in AI-Driven Environments*

Table I outlines the definitions of the five key constructs examined in this study: Data Ownership and Control (DOC), Parental Data Sharing (PDS), Perceived Risks and Benefits (PRB), Transparency and Trust (TT), and Education and Awareness (EA). Privacy in AI systems is influenced by the interplay of user agency, system transparency, and external factors [10]. Young digital citizens interact with AI applications that continuously collect, analyze, and process their data, yet they often lack the knowledge or authority to effectively manage their digital footprints [11]. While AI systems offer personalized experiences and enhanced user engagement, they also raise concerns about data ownership, trust, and perceived risks [5], [12]. The complexity of AI-related privacy challenges calls for a structured comparative analysis of how different stakeholder groups (youth, parents and/or educators, and AI professionals) conceptualize privacy, navigate trade-offs, and evaluate transparency concerns within AI-driven ecosystems.

A significant challenge lies in the power imbalance surrounding data ownership and management. Young users are expected to interact with AI systems but are rarely afforded meaningful control over their personal information [4]. The cryptic nature of AI decision-making and algorithmic profiling exacerbate this issue, limiting users' ability to opt out or influence how their data is collected and used [1]. Parental mediation further complicates matters, as efforts to protect youth may inadvertently lead to the oversharing of personal information, especially in AI-based family applications [13]. Research suggests that intergenerational privacy expectations often differ, with parents prioritizing security while younger individuals want more autonomy and flexibility in managing their data [14].

TABLE I. CONSTRUCTS AND DEFINITIONS

| Construct | Definition |
|---|---|
| Data Ownership and Control (DOC) | It is the degree to which young people have control over their personal data and engage in discussions about privacy. |

| Construct | Definition |
|---|---|
| Parental Data Sharing (PDS) | It is the degree to which parents exercise their rights to share children's data and consider the implications of doing so. |
| Perceived Risks and Benefits (PRB) | It is the degree to which individuals perceive risks, ethical concerns, and benefits related to the use of personal data by AI systems. |
| Transparency and Trust (TT) | It is the degree to which transparency in data usage influences trust in AI systems. |
| Education and Awareness (EA) | It is the degree to which stakeholders are informed about privacy and ethical issues associated with AI. |

Furthermore, the balance between privacy risks and perceived benefits significantly affects how young people interact with AI. Many young users recognize the advantages of personalized recommendations, social connectivity, and AI-enhanced learning tools, but often underestimate the long-term privacy ramifications of sharing their data [15], [16]. This aligns with the privacy paradox, where stakeholders express concerns about data exploitation while also participating in high-disclosure behaviours due to immediate perceived benefits [12]. Transparency is crucial in influencing these behaviours, as AI systems that clearly communicate data practices and provide accessible privacy settings are more likely to build trust and encourage responsible engagement [6]. Nevertheless, research suggests that most AI-driven platforms function with limited transparency, making it difficult for young users to evaluate who has access to their data and how it's being used [7].

Improving AI literacy and digital awareness is also essential for equipping young digital citizens to navigate these challenges effectively. Studies show that adolescents who receive structured privacy education and ethical AI training have enhanced data protection practices and risk assessment skills [8], [13]. However, current privacy education initiatives often fall short, as many programs fail to address the algorithmic intricacies and ethical dilemmas associated with AI-driven environments [5]. Interactive learning efforts and peer-led AI literacy programs have shown potential in bridging this gap, empowering young users with practical strategies to manage privacy settings, evaluate data-sharing trade-offs, and critically assess AI-generated content [14].

This research moves beyond the traditional discourse on privacy and AI ethics by investigating how young digital citizens actively negotiate privacy boundaries in practical AI interactions. The findings contribute to the development of youth-centric privacy policies, ethical AI frameworks, and regulatory measures that prioritize autonomy, transparency, and digital resilience.

### *C. Transparency, Trust, and User Autonomy in AI Privacy Through a Grounded Theory Perspective*

Transparency and trust are integral to privacy management in AI-driven environments, particularly for young digital citizens who navigate cryptic algorithmic systems with limited agency [11]. While AI applications offer personalization and efficiency, their lack of explainability and user control has fuelled skepticism about data collection, processing, and sharing practices [6]. Traditional research has treated transparency as a static variable, but a grounded theory approach enables a dynamic understanding of how young users conceptualize and negotiate trust in AI ecosystems [7]. By conducting a comparative analysis of stakeholder responses, this study identifies key discrepancies in trust formation, transparency expectations, and user autonomy between young digital citizens, parents and/or educators, and AI professionals, offering a structured evaluation of their privacy concerns [5].

Research indicates that trust in AI is influenced by contextual factors, with young users demonstrating higher trust in AI-powered tools when transparency aligns with their expectations of data control and ethical safeguards [12][13]. AI-driven applications in education and healthcare, for instance, demand the highest levels of transparency, as young users and stakeholders perceive privacy risks in these domains as particularly significant [17]. In contrast, social media and entertainment platforms often operate under more lenient transparency expectations, despite similarly extensive data collection practices [14]. This discrepancy highlights that transparency concerns among youth are not uniform, reinforcing the need for context-specific AI governance models rather than one-size-fits-all regulatory approaches [18].

### *D. Multi-Stakeholder Privacy Negotiation in AI Systems: An Analytical Perspective*

Privacy governance within AI-driven environments is not a top-down process but an evolving negotiation among multiple stakeholders. While grounded theory analysis provides a framework for identifying emerging privacy themes, privacy expectations do not operate in isolation; rather, they are shaped through continuous interactions between young digital citizens, parents, educators, and AI professionals. These stakeholders hold divergent views on privacy autonomy, transparency requirements, and regulatory oversight, leading to negotiated trade-offs that influence the structure of AI governance [19].

Recent research highlights how privacy decision-making is often influenced by multiple contextual factors, including technological literacy, regulatory frameworks, and parental involvement [20], [21]. Similarly, prior studies suggest that youth privacy behaviors are often driven by risk-benefit analysis, reflecting the need to assess AI governance through a comparative analytical lens rather than relying on static regulatory models [22]. While existing models explore privacy through theoretical frameworks, they often fail to capture the dynamic interplay between stakeholder priorities in AI-driven ecosystems.

To address these complexities, this study introduces a stakeholder-driven negotiation model, conceptualizing privacy governance as an interactive process shaped by competing priorities, ethical concerns, and governance expectations. Unlike static privacy frameworks, this model examines privacy tensions across diverse user groups, recognizing that youth, parents, educators, and AI professionals negotiate privacy in real-time interactions rather than through fixed policy structures. By framing privacy governance as a dynamic negotiation process, this model offers a comparative framework to assess how stakeholders engage with AI transparency, data control mechanisms, and risk evaluations. The following section explores key research gaps, emphasizing why a negotiation-based perspective is crucial for developing adaptive, inclusive AI governance models that align with real-world privacy concerns.

### *E. Bridging Research Gaps: Toward a Youth-Centric AI Privacy Framework*

Despite growing awareness of AI-related privacy threats, current research remains disjointed in its exploration of how young digital citizens participate in privacy decision-making [23]. Many studies focus primarily on adult privacy behaviors or regulation compliance, often overlooking the unique socio-cognitive elements that influence youth privacy perspectives within AI systems [13]. This gap is particularly evident in the absence of a structured comparative framework to evaluate how privacy concerns evolve differently across stakeholder groups, reflecting variations in risk perception, transparency expectations, and control mechanisms[15]. Existing AI governance models frequently overlook the agency of young users, treating them as passive beneficiaries of privacy protection rather than active participants in the conceptualization of AI governance [5]. Additionally, while digital literacy programs address privacy issues, they seldom provide systematic approaches to help young people navigate the trade-offs between the benefits of personalization and the risks to data security [11]. These gaps highlight the need for a theoretical framework that captures the complexities of youth privacy management in AI-driven environments.

This research addresses these shortcomings by using grounded theory to provide a youth-centric AI privacy framework, offering a dynamic, empirical understanding of how young users perceive, navigate, and respond to privacy challenges in AI contexts [7]. This approach enables the detection of emerging privacy behaviours influenced by AI transparency, parental mediation, and trust dynamics, rather than relying on static privacy models [6]. This study also advances the dialogue by advocating for adaptive privacy measures, participatory AI governance, and policy frameworks that integrate youth perspectives into AI design and legislation [12]. Furthermore, it highlights the development of interactive AI literacy initiatives that go beyond basic knowledge to equip young digital citizens with practical resources for safeguarding their privacy [14]. By prioritizing youth agency and stakeholder collaborations, this research lays the groundwork for ethical AI regulations and privacy solutions that align with the evolving expectations and opinions of young users.

## III. Methodology

### *A. Research Goal and Questions*

The primary objective of this study is to explore how young digital citizens and key stakeholders, parents, educators, and AI professionals, navigate privacy concerns in AI-driven environments. Using grounded theory [9], this research examines emerging themes related to data ownership, trust, transparency, parental influence, and AI literacy. This study takes an inductive approach to identify how these factors shape youth privacy behaviors. The study is guided by five key research questions:

- **RQ1:** How do young digital citizens, parents/educators, and AI professionals perceive privacy risks and responsibilities in AI-driven ecosystems?
- **RQ2:** What are the key factors influencing data ownership, user control, and privacy decision-making among different stakeholder groups in AI environments?
- **RQ3:** How do varying levels of AI literacy and digital awareness impact privacy management behaviors and attitudes toward transparency?
- **RQ4:** What role does stakeholder collaboration (youth, parents/educators, and AI professionals) play in shaping effective AI privacy governance frameworks?
- **RQ5:** How can participatory design approaches and adaptive privacy policies improve AI transparency, trust, and ethical AI system development?

These variables, informed by the research questions, converge to provide actionable insights for Ethical AI Development. The definitions of these constructs are outlined in Table I, which details their scope and focus within the study.

### *B. Research approach and analytical framework:*

This study employs a grounded theory approach to explore how young digital citizens, parents and/or educators, and AI professionals perceive and navigate privacy concerns in AI-driven environments. Grounded theory was applied as an inductive analytical method, allowing themes to emerge organically from the data rather than being constrained by pre-existing theoretical frameworks. Through a detailed analysis of both quantitative and qualitative responses, recurring privacy patterns were identified across stakeholder groups. These patterns were systematically categorized into thirteen generalized privacy themes, including data control importance, perceived data control, comfort with data sharing, parental data sharing, parental data rights, AI privacy concerns, perceived data benefits, data usage transparency, transparency perception, system data trust, privacy protection knowledge, digital privacy education, and AI privacy awareness. These themes formed the foundation for the development of the Privacy-Ethics Alignment in AI (PEA-AI) Model.

The study was structured around five core privacy constructs-Data Ownership and Control (DOC), Parental Data Sharing (PDS), Perceived Risks and Benefits (PRB), Transparency and Trust (TT), and Education and Awareness (EA)-which initially guided data collection. However, rather than assuming fixed relationships between these constructs, a thematic analysis was conducted at the item level to examine how specific survey responses reflected variations in privacy concerns, risk perceptions, and governance expectations across different stakeholder groups. The survey responses were analyzed by calculating mean values for each item across young digital citizens, parents and/or educators, and AI professionals, providing a structured way to compare privacy attitudes and identify patterns in stakeholder perspectives. These findings were further contextualized through qualitative insights from open-ended responses, interviews, and focus groups, ensuring that the emerging privacy themes were grounded in real-world stakeholder concerns rather than predefined categories.

Through this iterative, comparative process, emerging themes were synthesized into broader stakeholder-driven privacy categories, allowing for a structured understanding of privacy dynamics. The thematic analysis revealed distinct stakeholder perspectives on data control, parental mediation, AI risks, transparency, and digital literacy, highlighting key differences in how each group conceptualizes privacy responsibilities. Based on these comparative insights, the PEA-AI Model was developed to provide a structured framework for analysing how privacy governance evolves through stakeholder engagement, disagreements, and negotiated privacy boundaries. Rather than treating privacy as a static concept, the model reflects how trust, transparency, and data control expectations shift based on stakeholder interactions and the broader ethical discourse surrounding AI systems.

The model is introduced in the Results section, where it serves as an analytical tool to compare stakeholder-driven privacy dynamics and ethical AI governance strategies. By organizing stakeholder responses into generalized privacy themes, the study offers a comparative framework that highlights key tensions, alignments, and gaps in privacy governance, ensuring that AI policies and design principles reflect diverse perspectives. This approach ensures that privacy governance is understood as an evolving discourse shaped by real-world stakeholder interactions, fostering a more adaptive, inclusive, and user-cantered understanding of AI ethics rather than relying on rigid theoretical assumptions.

### *C. Research Design*

The present study received ethics approval from the Vancouver Island University Research Ethics Board (VIU-REB). The approval reference number #103116 was given for behavioral application/amendment forms, consent forms, interview and focus group scripts, and questionnaires. An initial pilot study was conducted with 6 participants, including members of the empirical research specialists from the University of Saskatchewan and Vancouver Island University. The pilot study aimed to evaluate the feasibility and duration of the research approach while refining the study design. Participants provided general feedback on the questionnaire, which guided the modification and restructuring of the final survey. The revised research model was then tested by collecting survey data. Survey data was collected by recruiting participants through flyers, personal networks, emails, and social networking sites, LinkedIn and Reddit. To reach the targeted youth demographic, several Vancouver Island school districts were contacted to assist in distributing the survey to their high-school students. Participation in the study was entirely voluntary and uncompensated. Participants were required to read and accept a consent form before starting the questionnaire, indicating their understanding of the study conditions outlined in the form. Online surveys were conducted through Microsoft Forms, with participants responding based on three designated demographics: AI Researchers and Developers, Parents and Teachers, and Young Digital Citizens (aged 16-19).

In addition to the questionnaires, interviews and focus groups were conducted with AI professionals, parents, and educators. A section of the questionnaire invited participants to provide their email addresses if they were interested in participating in interviews and/or focus groups. After contacting consented participants, 12 interviews and 2 focus groups were conducted: one with 4 AI professionals and another with 5 parents and/or educators. Before the interviews and focus groups, all participants reviewed and accepted a consent form. Sessions were conducted and transcribed using Microsoft Teams, with participants instructed to keep their videos off to ensure anonymity. Additionally, OpenAI's ChatGPT was utilized for grammatical and language refinement to ensure clarity and coherence in the research writing process [24].

The survey instruments were adapted from constructs validated in prior studies [25], [26], [27], [28], [29], [30], [31], [32], [33], [34], [35], [36], [37], [38]. The instruments consist of 3 indicators for Data Ownership and Control (DOC), 2 indicators for Parental Data Sharing (PDS), 4 indicators for Perceived Risk and Benefits (PRB), 3 indicators for Trust and Transparency (TT), 3 indicators for Education and Awareness (EA), and 3 open-ended discussion questions. The items (questions) within these constructs are outlined in Table II. Survey responses were measured on a 5-point Likert scale, with most items used for quantitative analysis. Notably, to ensure consistency in outcomes, we reversed the scale for items in PRB for AI professionals and swapped items 1 and 2 in PDS for educators/parents to algin contextually with the items for other demographics. Qualitative analysis utilized open-ended questions, two indicators from PRB, interview responses, and focus group discussions.

The following naming conventions were used for qualitative responses: survey participants were labeled as (S-YDC #*X*) for young digital citizens, (S-PE #*X*) for parents and educators, and (S-AIP #*X*) for AI Professionals. Interview participants were referred to as (I-[Role] #*X*), specifying their role (e.g., I-Parent #1 or I-Educator #2). For focus group participants, we use a group identifier and role, such as (FG1-Educator #3).

### *D. Participants demographics*

Out of 482 participants, 461 completed the survey questionnaire: 176 young digital citizens (aged 16–19), 132 parents and/or educators, and 153 AI professionals. After data cleaning, we retained 127 valid responses from educators and/or parents, 146 from AI professionals, and 151 from young digital citizens for analysis. Of the 127 valid responses from educators and/or parents, 54 identified as parents, 46 identified as educators, and 28 identified as both. Among the 146 valid responses from AI professionals, 46 identified as AI developers, 98 as AI researchers, and 2 as both. Twelve interviews were conducted, with 9 interviewees identifying as a parent and/or educators and 3 as AI professionals. Two focus groups were conducted, with 4 participants identifying as AI professionals and 5 as parents and/or educators. Table III highlights the demographic characteristics of the participants.

TABLE II. CONSTRUCTS AND ITEMS

| Construct | Items |
|---|---|
| Data Ownership and Control (DOC) | doc1: Importance of users having control over their personal data.<br>doc2: Frequency of considering user data control in work.<br>doc3: Feasibility/comfortability of implementing data control mechanisms. |
| Parental Data Sharing (PDS) | pds1: Handling data shared by parents on behalf of children.<br>pds2: Importance of obtaining consent from young users. |
| Perceived Risks and Benefits (PRB) | prb1: Concern about ethical/privacy implications<br>prb2: Significance of benefits in justifying data use.<br>Qpen-Ended Question: Primary risks associated with personal data use.<br>Qpen-Ended Question: Benefits AI systems provide by using personal data. |
| Transparency and Trust (TT) | tt1: Importance of transparency about data usage.<br>tt2: Perception of transparency in current AI systems.<br>tt3: Belief that increasing transparency improves user trust. |
| Education and Awareness (EA) | ea1: Knowledge about privacy issues related to AI systems.<br>ea2: Belief that users receive adequate training on privacy.<br>ea3: Importance of being educated on privacy and ethical issues/ Adequacy of privacy information. |

TABLE III. PARTICIPANT'S DEMOGRAPHICS

| Respondents' characteristics | Percentage(%) | | | Number of participants (n) | | |
|---|---|---|---|---|---|---|
| | Survey | Interview | Focus Group | Survey | Interview | Focus Group |
| **Young Digital Citizens** | 35.6 | 0.0 | 0.0 | 151 | 0 | 0 |
| **Parents** | 12.7 | 8.3 | 11.1 | 54 | 1 | 1 |
| **Educators** | 10.8 | 41.7 | 33.3 | 46 | 5 | 3 |
| **Both Parent and Educator** | 6.4 | 25.0 | 11.1 | 27 | 3 | 1 |
| **AI Developers** | 10.8 | 16.7 | 33.3 | 46 | 2 | 3 |
| **AI Researchers** | 23.1 | 8.3 | 11.1 | 98 | 1 | 1 |
| **Both AI Developers and Researcher** | 0.5 | 0.0 | 0.0 | 2 | 0 | 0 |

## IV. RESULTS & ANALYSIS

This study expands upon our previous research by incorporating additional responses from young digital citizens, allowing for a more robust and comprehensive analysis of their perspectives on privacy in AI systems. This study employs the Privacy-Ethics Alignment in AI (PEA-AI) Model as an analytical framework to examine how privacy expectations, transparency concerns, and AI governance strategies vary across stakeholder groups. Derived through a comparative thematic analysis of stakeholder responses, the model offers a structured approach to understanding how young digital citizens, parents and/or educators, and AI professionals negotiate privacy in AI-driven environments. Rather than establishing direct causal relationships, the model identifies key tensions, alignments, and differences in privacy perceptions across the five core constructs: Data Ownership and Control (DOC), Parental Data Sharing (PDS), Perceived Risks and Benefits (PDS), Transparency and Trust (TT), And Education and Awareness (EA).

This study employs a dual-layered analytical approach to explore stakeholder privacy concerns in the AI ecosystem, combining quantitative and qualitative methods to uncover critical privacy negotiation points and demonstrate how multi-stakeholder discourse shapes privacy governance.

For quantitative analysis, Microsoft Excel was used to structure, clean, and manage the collected survey data, ensuring consistency in the dataset. Descriptive statistical analysis was conducted across four distinct groups: young digital citizens, parents and educators, AI developers and researchers, and a combined dataset consolidating all responses. This analysis allowed for a structured comparison of responses, revealing trends in data control, transparency concerns, perceived risks, and awareness levels. Mean values were calculated and sentiment levels were categorized accordingly for key constructs: DOC, PDS, PRB, TT, and EA. By systematically comparing these constructs, we identified how different stakeholder groups perceive and prioritize privacy-related concerns, illustrating the thematic contrasts in their expectations and decision-making processes. This approach ensures the analysis evaluates the interplay of privacy constructs within and across groups, rather than examining them in isolation.

To contextualize these findings, qualitative data from open-ended responses, interviews, and focus group discussions were analyzed thematically. This revealed common patterns in participant concerns, perceptions, and recommendations. Together, these methods offer a comprehensive understanding of privacy concerns and values, allowing for a deeper exploration of stakeholder priorities, knowledge gaps, and expectations from AI privacy frameworks.

### A. *Descriptive Statistics*

Our quantitative survey used a 5-point Likert scale to compare mean responses across five key constructs: Data Ownership and Control (DOC), Parental Data Sharing (PDS), Perceived Risks and Benefits (PRB), Transparency and Trust (TT), and Education and Awareness (EA). The mean scores for each construct varied across three key demographics-young digital citizens, parents and/or educators, and AI professionals. The results are visually represented in Fig. 1 (heatmap) and detailed in Table IV (comparative analysis). Overall, while adults agree on user control, transparency, and engagement with AI, youth show lower trust and awareness, highlighting the need for targeted interventions to bridge this gap.

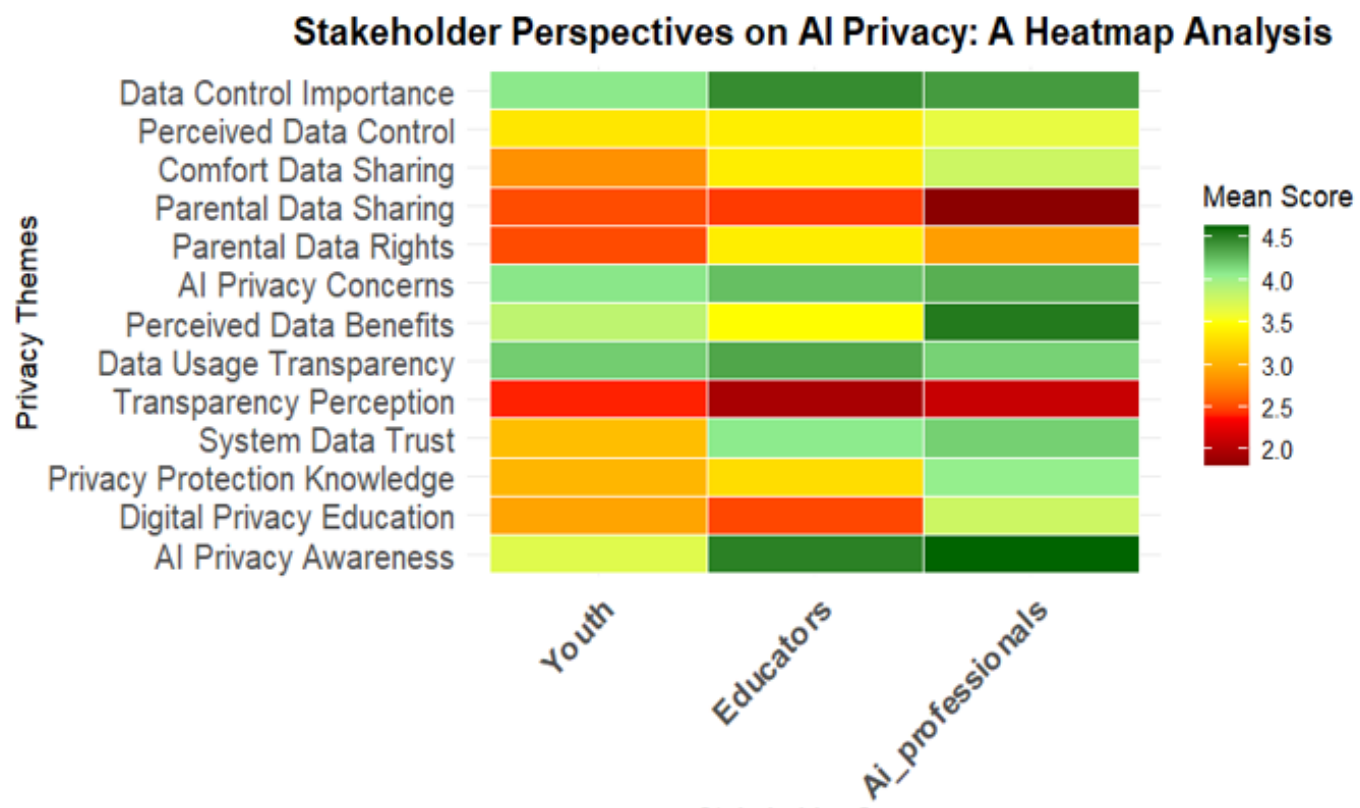

*Fig. 1. Heatmap Analysis*

TABLE IV. COMPARATIVE ANALYSIS (MEAN SCORE)

| Theme | Youth | Parent / Educator | AI Professionals |
|---|---|---|---|
| Data Control Importance | 4.09 | 4.46 | 4.40 |
| Perceived Data Control | 3.35 | 3.40 | 3.64 |
| Comfort Data Sharing | 2.83 | 3.40 | 3.79 |
| Parental Data Sharing | 2.52 | 2.46 | 1.82 |
| Parental Data Rights | 2.52 | 3.39 | 2.90 |
| AI Privacy Concerns | 4.09 | 4.25 | 4.32 |
| Perceived Data Benefits | 3.87 | 3.50 | 4.53 |
| Data Usage Transparency | 4.19 | 4.35 | 4.17 |
| Transparency Perception | 2.41 | 1.96 | 2.12 |
| System Data Trust | 3.09 | 4.08 | 4.18 |
| Privacy Protection Knowledge | 3.05 | 3.29 | 4.05 |
| Digital Privacy Education | 2.93 | 2.50 | 3.79 |
| AI Privacy Awareness | 3.69 | 4.50 | 4.63 |

### *B. Comparative Analysis of AI Privacy Constructs Across Stakeholder Groups*

The comparative analysis of privacy constructs across young digital citizens, parents and/or educators, and AI professionals reveals distinct variations in how different stakeholders perceive and engage with AI-driven privacy concerns. By examining mean scores across thematic categories, key differences emerge in the conceptualization of data control, parental data sharing, perceived risks and benefits, transparency, trust, and education.

#### *1) Data Ownership and Control (DOC)*

*a) Data Control Importance:* Parents/Educators rated this highest (4.46), followed by AI Professionals (4.39) and Youth (4.08). This suggests that adult stakeholders, especially educators and researchers, strongly advocate for user control over personal data, reinforcing its role in ethical AI development.

*b) Perceived Data Control:* AI Professionals (3.64) reported feeling more in control over their data compared to Parents/Educators (3.40) and Youth (3.35). The relatively lower score among youth suggests a potential gap in privacy self-efficacy, necessitating better user-centric privacy mechanisms.

*c) Comfort with Data Sharing:* AI Professionals (3.79) displayed the highest comfort in sharing personal data, followed by Parents/Educators (3.39), with Youth reporting the lowest comfort (2.83). This reflects a generational divide in risk perception, with youth demonstrating greater apprehension towards personal data disclosure.

#### *2) Parental Data Sharing (PDS)*

*a) Parental Data Sharing Practices:* AI Professionals reported the lowest support for parental data sharing (1.81), followed by Parents/Educators (2.46), and Youth (2.52). These relatively low scores indicate widespread concerns about the appropriateness of parental involvement in youth data decisions.

*b) Parental Data Rights:* Parents/Educators (3.39) rated parental data rights the highest, while AI Professionals (2.90) and Youth (2.51) expressed lower confidence in this construct. Notably, AI professionals, favored youth consent mechanisms, prioritizing autonomy over parental governance in data-related decisions.

#### *3) Perceived Risks and Benefits (PRB)*

*a) AI Privacy Concerns:* All three groups expressed strong privacy concerns, with AI Professionals scoring the highest (4.31), followed by Parents/Educators (4.25) and Youth (4.09). This consensus highlights the universal recognition of ethical challenges posed by AI data governance.

*b) Perceived Data Benefits:* AI Professionals (4.53) rated data benefits significantly higher than both Youth (3.86) and Parents/Educators (3.50). These findings suggest that while AI professionals see tangible advantages in data-driven AI advancements, youth and educators remain more cautious, reflecting a trust gap in AI benefit perception.

#### *4) Transparency and Trust (TT)*

*a) Data Usage Transparency:* Transparency was considered highly important across all stakeholder groups, with Parents/Educators scoring highest (4.34), followed by Youth (4.19) and AI Professionals (4.17). This reinforces the demand for increased transparency mechanisms in AI governance.

*b) Transparency Perception:* Despite valuing transparency, stakeholders perceived existing AI transparency measures as insufficient. Parents/Educators rated transparency perception lowest (1.96), followed by AI Professionals (2.11), and Youth (2.41). These results indicate a strong disparity between their expectations and the current implementation of transparency in AI systems.

*c) System Data Trust:* AI Professionals (4.18), followed by Parents/Educators (4.07), exhibited relatively higher trust in AI systems, while Youth expressed significantly lower trust (3.09). These findings suggest that youth are more skeptical of AI governance practices, reinforcing the necessity of improved explainability measures.

### 5) Education and Awareness (EA)

*a) Privacy Protection Knowledge:* AI Professionals reported the highest levels of privacy knowledge (4.04), followed by Parents/Educators (3.29) and Youth (3.04). The significant gap between professionals and youth suggests an urgent need for targeted AI privacy education initiatives.

*b) Digital Privacy Education*: AI Professionals rated privacy education substantially higher (3.79) compared to Youth (2.93) and Parents/Educators (2.50). These results highlight a potential divide in AI literacy, where non-technical stakeholders may lack the resources to fully understand privacy frameworks.

*c) AI Privacy Awareness:* All groups strongly agreed on the importance of AI privacy education, with AI Professionals rating it highest (4.63), followed by Parents/Educators (4.50) and Youth (3.68). The widespread alignment in this area suggests broad recognition of the need for continuous privacy education programs.

## C. Qualitative findings

In addition to the quantitative findings, the qualitative data revealed key stakeholder tensions regarding privacy governance in AI-driven environments. A significant divide emerged between young digital citizens and parents/educators, particularly in the realm of parental consent and data-sharing authority [39]. While youth participants expressed frustration over their lack of control and transparency in how their data is handled, many parents and educators viewed youth privacy as something that should be actively managed rather than autonomously controlled. One young respondent voiced their concern, stating, "I am mainly concerned about what data is being taken and how it is used, as I feel we often aren't informed clearly about what data is being taken and used." (S-YDC #5). Conversely, an educator emphasized the role of awareness rather than outright control, explaining, "Many children and adolescents will use AI without considering their own privacy (similar to how many use social media). A lack of education regarding the risks of sharing personal information on the internet can lead to students potentially misusing AI." (S-PE #40). This tension highlights a fundamental gap between youth demands for autonomy and parental concerns about informed decision-making in AI privacy governance.

Another point of contention emerged between educators and AI professionals regarding the trustworthiness of AI applications. Educators largely expressed skepticism over whether AI systems genuinely safeguard personal data, citing concerns about long-term data retention, algorithmic biases, and lack of oversight in AI-driven decision-making. One educator articulated this skepticism, stating, "I do not trust that information gathered by AI will be used presently or in the future in an informed manner for the benefit of the individual, but rather fear its exploitation on both an individual and mass level." (S-PE #10). However, AI professionals generally framed privacy risks as technical challenges that could be addressed through improved security measures rather than as inherent flaws in AI systems. An AI researcher highlighted the difficulty in tracing data flows, stating, "Once data goes into an AI system, it's tough to know where it ends up or who else can see it." (S-AIP #45). This divergence suggests that while educators advocate for broader regulatory oversight and ethical considerations, AI professionals emphasize governance through internal safeguards and privacy-preserving technologies.

The issue of AI-driven surveillance and profiling further highlights the complex negotiation between young digital citizens and AI professionals. Youth participants expressed deep concern over AI's ability to track, categorize, and potentially manipulate their behaviours without their explicit consent. One respondent remarked, "I feel uncomfortable knowing AI can recognize my face in public places." (S-YDC #93), while another feared long-term profiling, stating, "AI will guess everything about us. Sensitive topics I research could be recorded forever." (S-YDC #84). AI professionals, however, tended to view these issues through the lens of data governance rather than direct surveillance. A researcher explained, "AI might accidentally recreate sensitive data from its training sets, exposing private information." (S-AIP #85). This suggests that while youth stakeholders perceive AI surveillance as an immediate privacy violation, AI professionals frame it as a solvable issue through stricter data management practices. These findings collectively highlight the necessity of a multi-stakeholder approach to AI privacy governance, one that not only strengthens technical safeguards but also considers the lived experiences of youth and the ethical concerns raised by educators and parents.

## D. The PEA-AI Model: The Negotiation Framework

The Privacy-Ethics Alignment in AI (PEA-AI) Model, presented in Fig. 2, provides an organized framework for understanding the development of privacy concerns and ethical considerations through multi-stakeholder engagement in AI governance. Using survey data and qualitative insights, this model compares and contrasts the privacy attitudes of three main stakeholder groups: young digital citizens, parents and educators, and AI professionals. Unlike traditional models that impose rigid privacy standards through prescriptive frameworks, the PEA-AI Model views the creation of ethical AI as an iterative negotiation process in which stakeholders' expectations of privacy and governance mechanisms evolve through interaction.

The PEA-AI Model introduces a multi-stakeholder privacy negotiation framework, where privacy constructs evolve through stakeholder engagement. This framework addresses four key tensions:

1) *Data Control vs. Trust:* Balancing youth autonomy with AI developers' risk-mitigation measures.

2) *Transparency vs. Perception:* Addressing the gap between AI's claimed transparency and user perception.

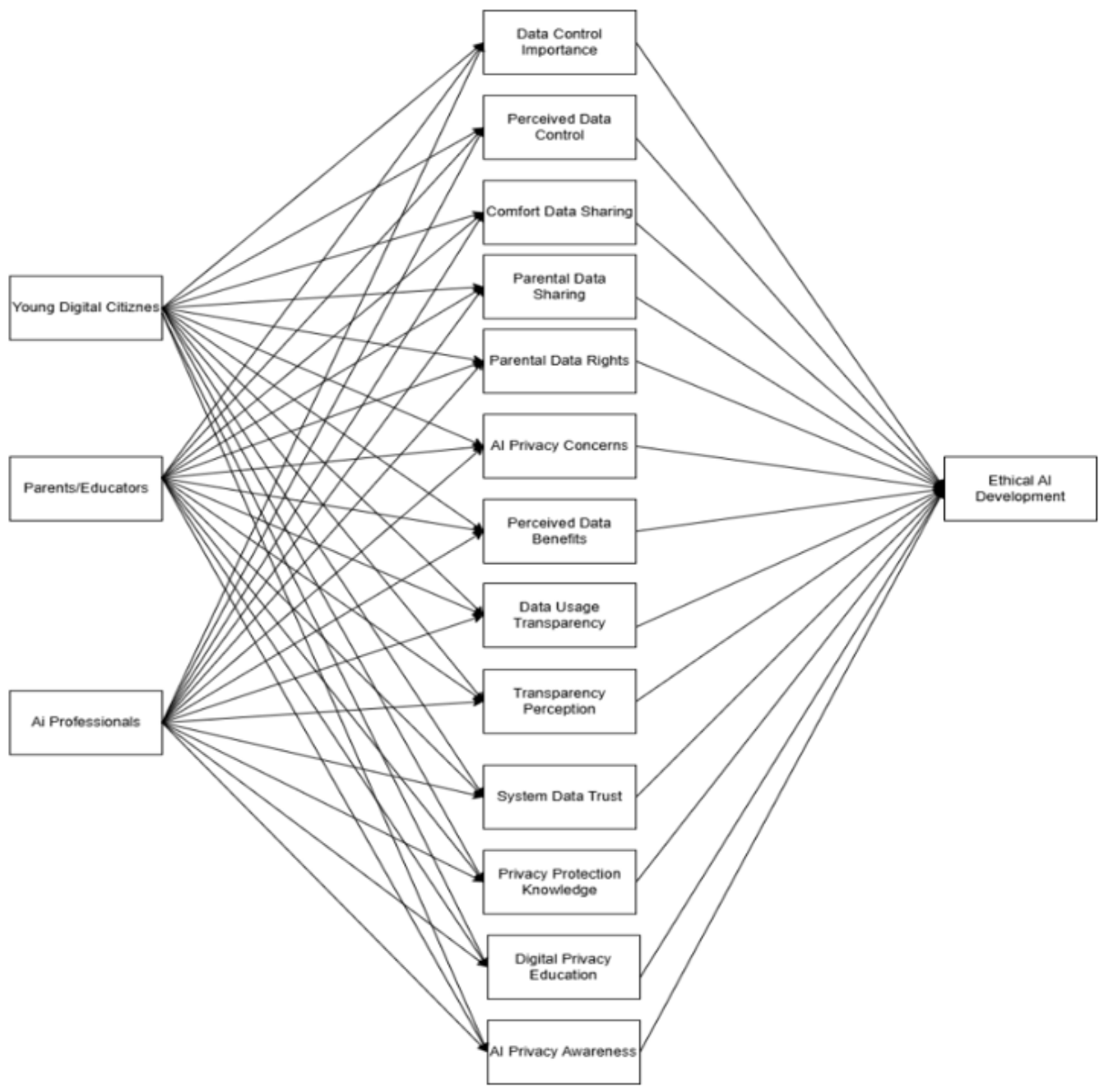

*Fig. 2. PEA-AI Model*

*3) Parental Rights vs. Youth Autonomy:* Negotiating consent mechanisms that respect youth agency while addressing parental concerns.

*4) Privacy Education vs. Awareness Deficit:* Strengthening digital literacy to enable informed AI interactions and empower users to navigate privacy challenges effectively.

This model was developed by systematically comparing responses to survey items concerning five fundamental privacy constructs: Data Ownership and Control (DOC), Parental Data Sharing (PDS), Perceived Risks and Benefits (PRB), Transparency and Trust (TT), and Education and Awareness (EA). By combining stakeholder responses into broad themes, the model focuses on how privacy perspectives influence governance and AI ethics rather than establishing fixed causal relationships.

As early adopters of AI, young people demonstrate a nuanced view of personal data protection. While they recognize the importance of data security, they often prioritize convenience over control in their privacy standards. Their responses suggest a willingness to allow AI-driven data collection, as the benefits of personalization, algorithmic recommendations, and social connectivity outweigh their privacy concerns. In contrast, educators and parents view privacy as a precaution, drawing attention to the risks associated with youth data disclosure and the need for regulatory oversight. Nevertheless, their perspectives reveal limited AI literacy, which hampers their ability to effectively guide youth in managing AI privacy settings. AI professionals, who are responsible for designing and implementing privacy safeguards, primarily view privacy through the lens of system performance and risk mitigation. While they acknowledge the challenges in achieving full transparency and user agency in AI-driven settings, their focus remains on compliance with existing regulations.

The PEA-AI Model highlights how stakeholder tensions shape privacy governance in the AI ecosystem. Despite all three categories engaging with AI systems, their expectations around data control, privacy, and trust often diverge. The comparative analysis reveals significant differences in privacy perceptions across themes. While all groups aim to control and manage data, their priorities vary: parents and educators prioritize security, young digital citizens value autonomy, and AI experts focus on technological limitations. Transparency and trust emerge as critical issues, with educators highlighting a lack of user-friendly disclosures, AI professionals acknowledging the challenges of comprehensive explainability, and young users demanding clarity and accessibility in AI systems.

By framing privacy as an ongoing negotiation rather than a static regulatory process, The PEA-AI Model advances AI development, stakeholder-driven governance, and policy interventions. It emphasizes the importance of aligning AI design with ethical considerations, ensuring that privacy measures reflect real user behaviour rather than hypothetical assumptions. The model advocates for the importance of multi-stakeholder involvement in governance, addressing the interests and constraints of various user groups. On the policy front, it guides adaptive regulation measures that balance the technical concerns of AI professionals, the privacy demand of young digital citizens, and the ethical considerations of parents and educators. Rather than prescribing a specific privacy solution, the model provides a comparative approach to understanding privacy conflicts in AI ecosystems, encouraging active participation from young digital citizens, educators, and AI professionals in the creation of ethical AI practices.

The PEA-AI Model makes several key contributions to the field of AI privacy governance. First, it introduces a stakeholder-driven approach to AI governance, unlike traditional models that focus solely on individual privacy attitudes. By integrating stakeholder negotiation, the model ensures that governance frameworks reflect the diverse needs and perspectives of all stakeholders, including young digital citizens, parents/educators, and AI professionals. Second, the model has significant policy implications, as it supports data protection frameworks that advocate for dual-consent mechanisms in AI data governance, balancing youth autonomy with parental oversight. Third, the model provides a foundation for AI design frameworks, enabling AI professionals to develop user-centered, privacy-enhancing technologies that prioritize transparency, accessibility, and ethical considerations. As AI evolves, the PEA-AI Model stresses the need for privacy governance to adapt to shifting stakeholder expectations, fostering a more inclusive and equitable digital society.

## V. Discussion

### A. *Privacy Perceptions in AI: Bridging Stakeholder Disparities*

The findings reveal significant disparities in how AI professionals, parents and/or educators, and young digital citizens perceive and manage privacy in AI ecosystems. While young digital citizens are increasingly aware of the privacy risks associated with artificial intelligence (AI), their understanding remains incomplete, which often leads to uneven privacy practices, particularly concerning data ownership, surveillance, and algorithmic profiling. Parents and educators, though advocating for protective measures and regulatory oversight, acknowledge their limited understanding of AI governance and emphasize the need for structural digital literacy programs. In contrast, AI professionals often focus on technological protections while overlooking the importance of user-centric openness and participatory governance. The misalignment highlights a critical gap: AI professionals prioritize system-level safeguards that may not correspond with user concerns, parents and educators call for more oversight but lack the expertise to implement effective interventions, and young users struggle with practical privacy management. These disparities highlight the importance of a cohesive, multi-stakeholder AI privacy framework that ensures privacy solutions are accessible, understandable, and sensitive to the real-life experiences of young digital citizens. Future AI governance initiatives should focus on participatory design approaches, fostering collaboration among youth, parents, educators, and AI professionals to create privacy policies that address the needs of all stakeholders.

### *B. Stakeholder Tensions in AI Privacy Management: Diverging Priorities and Overlapping Concerns*

The data-driven methodology of this study reveals significant knowledge gaps and control mechanisms, illustrating how different stakeholder groups comprehend and manage AI privacy. Privacy management is dynamic and context-dependent, contrasting with static privacy frameworks that assume consistent user behaviour. Young digital citizens expressed concern over data ownership but lacked the tools to exercise control, likely due to their limited awareness of data governance processes. In their advocacy to implement protective measures, parents and educators often misunderstand the intricacies of AI data flows, leading to adopting restrictive rather than empowering privacy policies. AI professionals, with their technical expertise and knowledge, tend to focus on system design and risk mitigation but overlook the practical challenges end-users confront in exercising privacy rights. These tensions reflect conflicting priorities: youth prioritize autonomy and transparency, educators and parents advocate for protective oversight, and AI professionals emphasize technical safeguards over user-centric governance. The PEA-AI model addresses these disparities by framing AI privacy governance as an evolving, multi-stakeholder negotiation, rather than a static regulatory structure. Addressing these conflicts requires co-developed mechanisms that promote transparency, user agency, and an adaptive governance framework responsive to diverse stakeholder concerns. This study contributes to the ongoing discussion on AI privacy governance by advocating for user-focused policies that integrate transparency, education, and participatory decision-making. Longitudinal studies tracking privacy attitudes over time could fill the gaps in our current understanding of privacy attitudes and inform the development of adaptive governance mechanisms for trustworthy AI systems.

### *C. The Privacy-Ethics Alignment in AI (PEA-AI) Model: A Stakeholder-Driven Framework for Ethical AI Governance*

The findings of this study informed the development of the Privacy-Ethics Alignment in AI (PEA-AI) Model, a stakeholder-driven framework that addresses the complexities of privacy governance in AI ecosystems. Rather than assuming fixed relationships between privacy constructs, this model conceptualizes ethical AI development as the outcome of an ongoing negotiation among young digital citizens, parents and/or educators, and AI professionals. The comparative analysis reveals that young users struggle with practical privacy management, parents and educators emphasize regulatory interventions, and AI professionals focus on system design and compliance-driven safeguards. These differences highlight the need for a dynamic, adaptable privacy governance framework that integrates technological feasibility with user-centric policies and educational initiatives. The PEA-AI model bridges these gaps by aligning privacy expectations, negotiating stakeholder responsibilities, and facilitating ethical AI decision-making'. Young digital citizens are often ill-equipped to exercise control over their personal data, making them vulnerable to cryptic data-gathering methods. Meanwhile, parents and educators prioritize protective measures but struggle to understand the technical complexities of AI privacy settings. AI professionals, focused on risk mitigation and compliance, may overlook the usability challenges everyday users face when navigating privacy concerns. These differences underscore the need for 'stakeholder-driven and integrated AI privacy management that balances technical, regulatory, and user-centric considerations. The PEA-AI addresses this need by acknowledging the multi-faceted nature of privacy concerns and advocating for flexible solutions that incorporate user agency, legislative protections, and technical practicality. Future AI governance must prioritize collaboration among educators, technologists, policymakers, and end-users, ensuring privacy safeguards are accessible, user-friendly, and in line with user expectations. Future studies should explore participatory design approaches, involving stakeholders in the development of AI privacy tools and policies, to achieve this goal.

### *D. Bridging the Gap Between Privacy Awareness and Practical Implementation*

While awareness of AI-related privacy risks is growing, significant gaps remain in translating this into practical, effective privacy management strategies. While young digital citizens understand the significance of data ownership and control, they face challenges when it comes to navigating the intricate settings of AI-powered platforms. Educators and parents understand the need for oversight but often lack the 'technical knowledge to help youth make informed privacy decisions. Meanwhile, AI professionals focus on developing privacy-enhancing technologies but may not prioritize the needs of non-technical users. The PEA-AI model underscores the need for bridge mechanisms that convert privacy knowledge into practical, user-friendly measures. This includes addressing gaps in literacy, designing intuitive privacy controls, and incorporating participatory governance models that adapt to user needs. Effective AI privacy solutions should simplify privacy controls, provide clear explanations of data usage, and offer real-time feedback to users. Multi-stakeholder educational initiatives, such as gamified AI ethics courses, youth-centric design concepts, and interactive privacy dashboards, can empower users to make informed decisions. Additionally, AI developers should conduct usability testing with diverse stakeholders to ensure that privacy measures are accessible and easy to comprehend. 'Future research should explore longitudinal methods to see if increased awareness leads to sustained privacy-protective, ensuring the privacy literacy translates into meaningful action.

### *E. The Role of Transparency in Building Trust and User Autonomy*

One of the most important aspects of AI privacy governance is transparency. However, our findings show that there is a disconnect between its perceived importance and its practical implementation. Users' faith in AI systems is eroded due to the absence of transparent, easily understandable explanations of data usage, even though all stakeholders agree that transparency is essential for building trust, users often lack clear, accessible explanations of how their data is collected, processed, and shared. Young digital citizens, in particular, express skepticism about AI-driven platforms due to the cryptic nature of algorithms, which makes it difficult to understand data flow. Educators and parents, though rightfully concerned about privacy, lack the knowledge to support stronger safeguards. AI professionals acknowledge the importance of openness but tend to put more emphasis on disclosures driven by compliance than on user-centric explainability, resulting in complicated privacy regulations that are difficult for users to understand. To address these challenges, AI governance must prioritize effective transparency measures, such as visual privacy indicators, interactive consent tools, and simplified explanations of AI processes tailored to different user demographics. Adaptive transparency models, which allow users to choose the level of insight they receive about AI data processing, can enhance user agency and decision-making. Future research should explore transparency-enhancing technologies (TETs), such as AI-generated privacy summaries or interactive data-flow diagrams, to empower users while maintaining system efficiency. Transparency must be a foundational factor of ethical AI design, ensuring accessibility, user agency, and trust.

### *F. Strengthening AI Privacy through User-Centric and Stakeholder-Inclusive Policy Interventions*

Policy interventions that are user-centric and balance technical advancements with ethical data practices are necessary for effective AI privacy control. While all stakeholder groups recognize the importance of privacy measures, there is often a disconnect between policy development and practical implementation. Young digital citizens face vulnerabilities due to untransparent permission processes, data commercialization, and algorithmic decision-making. Parents and educators advocate for more digital literacy programs but lack the knowledge and resources to adequately guide youth in 'navigating interactions with artificial intelligence. AI professionals tend to focus on privacy frameworks driven by compliance rather than user experience, which means they tend to overlook the usability challenges faced by non-technical users. To address these gaps, AI privacy policies need to be co-designed with direct stakeholder involvement, ensuring interventions are explicit, practical, and adaptable for different users. Regulations should mandate simplified privacy settings to enable young users and their guardians to manage their data effectively without needing technical expertise. Age-appropriate privacy standards, similar to child protection laws, should be expanded to address AI-specific risks, such as data-driven profiling and content manipulation. To further equip users to comprehend, evaluate, and manage privacy concerns, consistent digital literacy education should be introduced into national curricula. Beyond user education, corporate accountability and transparent enforcement mechanisms are essential to guarantee that AI systems adhere to ethical data governance standards. To ensure that privacy issues are detected and addressed before AI systems are deployed, it is imperative that these organizations conduct multi-stakeholder privacy impact studies. Additionally, advisory boards that include young digital citizens, educators, and AI professionals can provide real-time policy suggestions as privacy challenges evolve. Future studies should investigate the effects of user-driven policy design on AI trust and adoption, ensuring regulatory frameworks adapt to the needs and expectations of diverse stakeholders. Policy interventions that focus on users can promote responsible and transparent AI development while protecting users' privacy rights by closing the gap between legislative safeguards and their actual use.

### *G. Theoretical Implications for AI Privacy Governance*

This study contributes to the theoretical landscape of AI privacy governance by proposing a stakeholder-driven model that places privacy management in the context of an evolving process rather than a static regulatory construct. Traditional models often adopt a binary approach, emphasizing either user agency or regulatory enforcement, but our research highlights the importance of the interplay between young digital citizens, parents and/or educators, and AI professionals in developing privacy models. In contrast to these earlier models, our research shows that privacy behaviors in AI ecosystems are influenced by data literacy, perceived risks, and trust-building processes. This research builds on previous work to expand current conceptions of privacy by introducing three new characteristics of AI privacy frameworks: youth agency, participatory governance, and adaptive transparency. Our findings suggest that privacy behaviors among youth are shaped by contextual awareness, digital literacy gaps, and external factors like parental mediation and platform'. This complexity necessitates more nuanced theoretical models that account for the evolving interplay between privacy expectations, technological advancements, and government structures. Furthermore, the study highlights the importance of socio-technical privacy frameworks, highlighting that ethical AI design, algorithmic opacity, and corporate accountability are all systemic elements that must be considered when examining AI privacy. By integrating user engagement, explainability, and transparency into theories of AI governance, this research advocates for a more inclusive approach to privacy scholarship. Instead of relying on top-down legal frameworks, future theoretical breakthroughs could delve into privacy as a relational construct, where stakeholders co-negotiate the balance between data autonomy, security, and ethical monitoring. In addition, as knowledge about AI grows and regulations change, longitudinal research should be implemented to track people's views about privacy over time. Our research advances the discourse on AI governance by advocating for adaptive, user-empowered theoretical models that move beyond compliance-driven frameworks.

### *H. Limitations and Future Works*

While this study offers useful insights regarding AI privacy governance from a stakeholder-driven perspective, it has several limitations. First, the perspectives of policymakers, corporate AI developers, and non-technical adult users were not included, potentially overlooking diverse views on privacy and regulation. Expanding the sample size and including these demographics in future research might increase the findings' generalizability. Additionally, the cross-sectional design captures attitudes and behaviours at a single point in time, limiting our ability to observe privacy perceptions change in response to technological innovation, new regulations, or changes in AI literacy. Longitudinal studies, tracking stakeholder perspectives over time could provide an improved understanding of privacy adaptation in AI ecosystems. Moreover, using qualitative interviews and self-reported survey results has its limitations, as these methods introduce the risk of biases due to respondents' limited technical knowledge or social desirability. These findings could be further supported by experimental studies, behavioural data, or case studies of real-world AI privacy solutions. Additionally, while the PEA-AI Model provides a robust framework for organizing various privacy issues, testing it in other AI settings, such as healthcare, banking, and smart city projects, could reveal how privacy risks and governance requirements vary across different domains. Finally, as AI continues to rapidly evolve, future research should explore adaptive privacy frameworks that incorporate new privacy-enhancing technologies, changes in regulation, and participatory governance mechanisms. Co-designing AI privacy solutions with end-users will be critical to ensure privacy safeguards are user-centric, transparent, and in line with the changing expectations of digital citizens.

## VI. CONCLUSION

Drawing on insights from young digital citizens, parents, educators, and AI professionals, this study adopted a stakeholder-driven approach to explore the complexities of AI privacy governance. Using grounded theory, the research identified evolving privacy concerns and systemic barriers, such as gaps in digital literacy, regulatory inconsistencies, and deficits in algorithmic transparency, that hinder effective privacy management in AI-driven environments. To address these challenges, the study introduced the Privacy-Ethics Alignment in AI (PEA-AI) Model, a stakeholder-centric framework that integrates five key privacy constructs: Data Ownership and Control (DOC), Parental Data Sharing (PDS), Perceived Risks and Benefits (PRB), Transparency and Trust (TT), and Education and Awareness (EA). The findings highlight the critical role of privacy education (EA) in empowering individuals to make informed decisions, directly influencing data control (DOC) and risk assessment (PRB). Additionally, transparency and trust (TT) emerged as fundamental drivers of user confidence and engagement, reinforcing the need for accessible, explainable AI governance mechanisms. However, the study reveals tensions in privacy decision-making, particularly the restrictive nature of parental data sharing (PDS), which may limit youth autonomy in managing their personal data. Ethical AI development must therefore prioritize user-friendly privacy controls, participatory governance, and multi-stakeholder collaboration to create adaptive, inclusive privacy frameworks.

The PEA-AI Model lays a foundation for understanding stakeholder-driven ethical AI development, emphasizing the dynamic interplay between stakeholder concerns and AI governance. By offering a structured, evidence-based approach to ethical AI policymaking and system design, the model provides a roadmap for aligning privacy expectations with practical governance strategies. While this study contributes to theoretical discussions on AI privacy governance, future research should broaden its demographic scope, refine measurement constructs for greater reliability, and adopt longitudinal methodologies to track shifts in privacy attitudes over time. Furthermore, embedding participatory design strategies in AI system development and policymaking will be essential to ensuring that privacy safeguards remain flexible, transparent, and aligned with the evolving needs of digital citizens.

## ACKNOWLEDGMENT

This project has been funded by the Office of the Privacy Commissioner of Canada (OPC); the views expressed herein are those of the authors and do not necessarily reflect those of the OPC.